\documentclass[aps,%twocolumn
nofootinbib,amssymb,amsmath,amsthm,
showpacs,showkeys]{revtex4}
\usepackage{latexsym}
\usepackage{amsmath,amsthm,amssymb}
\newcommand{\bastar}{\begin{eqnarray*}}
\newcommand{\eastar}{\end{eqnarray*}}
\newskip\humongous \humongous=0pt plus 1000pt minus 1000pt

\newif\ifdtup

\relax
\newcommand{\be}{\begin{equation}}
\newcommand{\ee}{\end{equation}}
\newcommand{\bea}{\begin{eqnarray}}
\newcommand{\eea}{\end{eqnarray}}

\newcommand{\pro}{\partial}

\newcommand{\difrac}{\displaystyle\frac}
\newcommand{\ba}{\begin{array}}
\newcommand{\ea}{\end{array}}

\newcommand{\nn}{\nonumber}

\newcommand{\hna}{\hat\nabla}
\newcommand{\bna}{\bar\nabla}
\begin{document}
\title{Quantum Gravity Model in the framework of Weyl-Cartan geometry}
\bigskip
\author{B. S. Park}
\email{psychist@phya.snu.ac.kr}
\affiliation{College of Natural Sciences, Seoul National University, Seoul 151-742, Korea \\}
\author{D. G. Pak}
\email{dmipak@gmail.com}
\affiliation{ Institute of Applied Physics, Uzbekistan National University,
        Vuzgorodok, Tashkent, 100-174, Uzbekistan \\}

\begin{abstract}
We study the Weyl vector fields which can play an important role in quantum gravity. The metric
obtains its dynamical content after
dynamical symmetry breaking in the phase of the
effective Einstein gravity which is induced by quantum
Weyl corrections. In low energy regime with scalar field there is a relation between the Weyl vector fields and the torsion fields. If this condition is given to Weyl vector fields and torsions, then the Lagrangian becomes like Maxwell type.
\end{abstract}

%%04.60.-m Quantum gravity
%%04.62.+v Quantum field theory in curved spacetime
%%11.30.Qc Spontaneous and radiative symmetry breaking
%%11.30.Cp Lorentz and Poincare invariance
\maketitle

\section{Introduction}
 In the former paper\cite{generaltorsion} we propose a special $R^2$-type model of Lorentz gauge
gravity which admits a topological phase at classical level
and has non-trivial quantum dynamics of torsion.
The proposed model is minimal in a sense that only the
contortion possesses dynamic degrees of freedom
whereas the metric does not.
We demonstrate that the contortion
has six propagating modes with spins $J=(2;1;0;0)$, exactly the same number of
physical degrees of freedom the metric tensor has in general. In the present paper we will study further in the case of Weyl-Cartan geometry. In torsion free case the Gauss-Bonnet Lagrangian with the Weyl vector fields has negative kinetic energy term. But the Yang-Mills type Lagrangian has positive kinetic energy term. And we will show that Weyl vector field also can be a candidate for quantum gravity, and Einstein-Hilbert term can be induced by quantum corrections due to the condensation of Weyl vector fields. Next we consider the Palatini formalism with Einstein-Hilbert action. And in this case we will show that the Weyl vector fields become source of the torsions. But in this case Weyl vector fields and torsions are not dynamical fields. And if we give this special relation between torsions and Weyl vector fields to the general quadratic curvature Lagrangian, the Weyl vector fields become $U(1)$ vector fields.

%--------------------------------------------------------------
\section{The Yang-Mills type Weyl Gravity}\label{Weyl Gravity}
In Weyl's geometry, besides the general coordinate transformation of Einstein, we deals with the (Weyl) gauge transformation under which any length gets multiplied by a factor $e^{\Lambda(x)}$. So $ds' = e^{\Lambda(x)}ds$. A local tensor $T(x)$ which is transformed as $T'(x)=e^{J\Lambda(x)}T(x)$ is called a co-tensor of weight $J$. Since $ds^2 = g_{\mu\nu} dx^\mu dx^\nu $, $g_{\mu\nu}$ is a co-tensor of weight 2 under understanding that the $dx^\mu$ are not affected by Weyl transformation : $g'_{\mu\nu} = e^{2\Lambda(x)} g_{\mu\nu}$.\footnote{The signature of the metric is (-,+,+,+).}

Now with above settings let us define following co-covariant derivative
\be
D_\mu = \partial_\mu - \Gamma_\mu + fJ W_\mu \equiv \nabla_\mu + fJ W_\mu
\ee
where $\Gamma_\mu$ is the Weyl gauge invariant connections, $f$ is coupling constant, $J$ is the weight of tensor density and $W_\mu$ is the Weyl vector fields.
The metric and the Weyl vector field transform like following under the Weyl gauge transformation.
\bea
&g'_{\mu\nu} = e^{2\Lambda(x)} g_{\mu\nu} \\
&W'_\mu = W_\mu - \dfrac{1}{f} \partial_\mu \Lambda
\eea
The co-covariant derivative acts on co-scalar $S$ of weight $J$ like
\be
D_\mu S = \partial_\mu S + fJ W_\mu S \equiv \tilde{\partial}_\mu S.
\ee
For co-vector $V_\mu$,
\bea
D_\mu V_\nu &=&\tilde{\partial}_\mu V_\nu -\Gamma_{\mu\nu}^{~~\rho}V_\rho =\partial_\mu V_\nu - \Gamma_{\mu\nu}^{~~\rho}V_\rho + Jf W_\mu V_\nu \nn \\
&=& \partial_\mu V_\nu - \hat{\Gamma}_{\mu\nu}^{~~\rho}V_\rho - f\left(\delta_\nu ^\rho W_\mu + \delta^\rho_\mu W_\nu - g_{\mu\nu} W^\rho \right) V_\rho + Jf W_\mu V_\nu \label{de}
\eea
where $\hat\Gamma_{\mu\nu}^{~~\rho}=  \bar\Gamma_{\mu\nu}^{~~\rho}  + K_{\mu\nu}^{~~\rho}$ and $\bar\Gamma_{\mu\nu}^{~~\rho}$ is the Christoffel symbol and $K_{\mu\nu}^{~~\rho}$ is the contortion tensor.
Here we note that
\be
D_\mu g_{\nu\lambda} = \nabla_\mu g_{\nu\lambda} + fJ W_\mu g_{\nu\lambda}=\nabla_\mu g_{\nu\lambda} + 2f W_\mu g_{\nu\lambda} =0.
\ee
So $\nabla_\mu g_{\nu\lambda} = - 2f W_\mu g_{\nu\lambda}\equiv - Q_{\mu\nu\lambda}$ which means the non-metricity.

Now let $V^{(0)\mu}$ be a weight 0 vector and let $V^{\mu}=\sqrt{-g}^{J/4}V^{(0)\mu}$. Then we have
\bea
[D_\mu, D_\nu]V^\rho &=& [D_\mu, D_\nu](V^{(0)\rho}(\sqrt{-g})^{J/4}) \stackrel{\because D_\mu g_{\nu\lambda}=0}{=}(\sqrt{-g})^{J/4}[D_\mu, D_\nu]V^{(0)\rho}\nn\\
&&%\stackrel{}
{\nearrow}\overset{\because\Gamma_{\mu\nu}^{~~\lambda}, W_\mu \texttt{ has no weight!}}{}\nn\\
&=& (\sqrt{-g})^{J/4}[\nabla_\mu, \nabla_\nu]V^{(0)\rho}=(\sqrt{-g})^{J/4}\left\{R_{\mu\nu\lambda}^{~~~\rho}V^{(0)\lambda}-t_{\mu\nu}^{~~\sigma}\nabla_\sigma V^{(0)\rho}\right\} \nn\\
&=&\left\{R_{\mu\nu\lambda}^{~~~\rho}(\sqrt{-g})^{J/4}V^{(0)\lambda}-t_{\mu\nu}^{~~\sigma}D_\sigma\left(\sqrt{-g}^{J/4} V^{(0)\rho}\right)\right\}\nn\\
&=&R_{\mu\nu\lambda}^{~~~\rho}V^{\lambda}-t_{\mu\nu}^{~~\sigma}D_\sigma V^{\rho}
\eea
Therefore $D_\mu$ and $\nabla_\mu$ have the same curvature and torsion.
Now under the decomposition (\ref{de}) the curvature tensor is split into three parts
\be
R_{\mu\nu\lambda}^{~~~\rho}=\bar{R}_{\mu\nu\lambda}^{~~~\rho}+\tilde{R}_{\mu\nu\lambda}^{~~~\rho}+Q_{\mu\nu\lambda}^{~~~\rho}
\ee
where
\bea
\bar{R}_{\mu\nu\lambda}^{~~~\rho}&=&2\left\{ \partial_{[\mu}\bar{\Gamma}_{\nu]\lambda}^{~~\rho}+\bar{\Gamma}_{[\mu|\sigma|}^{~~\rho}\bar{\Gamma}_{\nu]\lambda}^{~~\sigma}\right\} \\
\tilde{R}_{\mu\nu\lambda}^{~~~\rho}&=& 2\left\{ \bna_{[\mu}K_{\nu]\lambda}^{~~\rho}+K_{[\mu|\sigma|}^{~~\rho}K_{\nu]\lambda}^{~~\sigma}
\right\} \\
Q_{\mu\nu\lambda}^{~~~\rho}
&=&2\left\{ \hna_{[\mu}\hat{Q}_{\nu]\lambda}^{~~\rho}+\hat{Q}_{[\mu|\sigma|}^{~~\rho}\hat{Q}_{\nu]\lambda}^{~~\sigma}
+\hat\Gamma_{[\mu\nu]}^{~~\sigma}\hat Q _{\sigma\lambda}^{~~\rho}\right\}\nn \\
&=&2\left\{ \bna_{[\mu}\hat{Q}_{\nu]\lambda}^{~~\rho}+\hat{Q}_{[\mu|\sigma|}^{~~\rho}\hat{Q}_{\nu]\lambda}^{~~\sigma}
+K_{[\mu|\sigma|}^{~~\rho}\hat Q _{\nu]\lambda}^{~~\sigma}+\hat Q_{[\mu|\sigma|}^{~~\rho} K_{\nu]\lambda}^{~~\sigma}\right\},
\eea
and
\bea
&&K_{\mu\nu}^{~~\rho}=K_{[\mu\nu]}^{~~\rho}+K_{(\mu\nu)}^{~~\rho}= \frac{1}{2}t_{\mu\nu}^{~~\rho}+t_{~(\mu\nu)}^{\rho}\\
&&\hat Q_{\mu\nu}^{~~\rho}= f\left(\delta_\nu ^\rho W_\mu + \delta^\rho_\mu W_\nu - g_{\mu\nu} W^\rho \right).
\eea
where $\bna_\mu$ is the covariant derivative containing only the Christoffel symbol part and $\hna_\mu$ is that containing both the Christoffel symbol part and the contortion tensor.

We can express the Weyl part curvature in terms of the Weyl vector fields and the torsion. Then,
\bea
Q_{\mu\nu\lambda\alpha}&=&2f g_{\lambda\alpha}\hat\nabla_{[\mu}W_{\nu]}-4fg_{[\alpha|[\mu}\hat\nabla_{\nu]|}W_{\lambda]}+4f^2 W_{[\mu}g_{\nu][\lambda}W_{\alpha]}-2f^2 W^\sigma W_\sigma g_{\alpha[\mu}g_{\nu]\lambda}\nn \\&&+f\left(t_{\mu\nu}^{~~\sigma}W_\sigma g_{\lambda\alpha} + t_{\mu\nu\alpha}W_\lambda -t_{\mu\nu\lambda}W_\alpha \right) \\
&=&2f g_{\lambda\alpha}\bna_{[\mu}W_{\nu]}-4fg_{[\alpha|[\mu}\bna_{\nu]|}W_{\lambda]}+4f^2 W_{[\mu}g_{\nu][\lambda}W_{\alpha]}-2f^2 W^\sigma W_\sigma g_{\alpha[\mu}g_{\nu]\lambda}\nn \\&&+f\left(4g_{[\alpha|[\mu}K_{\nu]|\lambda]}^{~~~\sigma}W_\sigma + t_{\mu\nu\alpha}W_\lambda -t_{\mu\nu\lambda}W_\alpha \right)
\eea
And
\bea
Q_{\mu\lambda} &=& g^{\nu\alpha} Q_{\mu\nu\lambda\alpha} \nn \\
&=& 2f\bar\nabla_\mu W_\lambda + 2f \bar\nabla_{[\mu}W_{\lambda]}+fg_{\mu\lambda}\bar\nabla_\alpha W^\alpha -2 f^2 W_\mu W_\lambda + 2 f^2 g_{\mu\lambda}W^\sigma W_\sigma \nn\\
&&-f\left(2K_{\mu\lambda}^{~~\sigma}W_\sigma  +t_{\mu~\lambda}^{~\alpha} W_\alpha -t_{\mu\alpha}^{~~\alpha}W_\lambda +g_{\mu\lambda}t^{\sigma~\alpha}_{~\alpha} W_\sigma \right) \\
\breve{Q}_{\mu\lambda}&\equiv& g^{\nu\alpha} Q_{\nu\mu\alpha\lambda} \nn \\
&=& 2f\bar\nabla_\mu W_\lambda - 2f \bar\nabla_{[\mu}W_{\lambda]}+fg_{\mu\lambda}\bar\nabla_\alpha W^\alpha -2 f^2 W_\mu W_\lambda + 2 f^2 g_{\mu\lambda}W^\sigma W_\sigma \nn\\
&&-f\left(2K_{\mu\lambda}^{~~\sigma}W_\sigma  +t_{\mu~\lambda}^{~\alpha} W_\alpha +t_{\mu\alpha}^{~~\alpha}W_\lambda +g_{\mu\lambda}t^{\sigma~\alpha}_{~\alpha} W_\sigma \right)
\eea
\bea
Q &=& g^{\mu\lambda}Q_{\mu\lambda}\nn \\
&=&6f \hna_\mu W^\mu + 6f^2 W^\mu W_\mu-2ft_{\mu\nu}^{~~\mu}W^\nu \nn \\
&=&6f \bna_\mu W^\mu +6fK_{\mu\nu}^{~~\mu} W^\nu + 6f^2 W^\mu W_\mu-2ft_{\mu\nu}^{~~\mu}W^\nu \nn \\
&=& 6 f \dfrac{1}{\sqrt{-g}} \partial_\mu \left(\sqrt{-g}W^\mu\right)+6f^2 W^\mu W_\mu +4ft_{\mu\nu}^{~~\mu}W^\nu \\
\breve{Q}&=& g^{\mu\lambda}\breve{Q}_{\mu\lambda}\nn \\
&=&6f \bna_\mu W^\mu  + 6f^2 W^\mu W_\mu+6ft_{\mu\nu}^{~~\mu}W^\nu.
\eea

Now for simplicity let us concentrate on the torsion free case. And we are going on computing the square terms of curvature. First the square of curvature tensors are
\bea
(R_{\mu\nu\lambda\alpha})^2 &=& (\bar R_{\mu\nu\lambda\alpha} + Q_{\mu\nu\lambda\alpha})(\bar R^{\mu\nu\lambda\alpha} + Q^{\mu\nu\lambda\alpha}) \nn \\
&=& \bar R_{\mu\nu\lambda\alpha}\bar R^{\mu\nu\lambda\alpha}+2 Q_{\mu\nu\lambda\alpha}\bar R^{\mu\nu\lambda\alpha} + Q_{\mu\nu\lambda\alpha} Q^{\mu\nu\lambda\alpha} \nn \\
&=& (\bar R_{\mu\nu\lambda\alpha})^2 + 8 f \bar R^{\mu\alpha} \bna_\mu W_\alpha \\
&&+4f^2 \left\{ 2(\bna_\mu W_\nu)^2 + 4(\bna_{[\mu} W_{\nu]})^2 + (\bna_\mu W^\mu)^2 - 2 \bar R ^{\mu\alpha} W_\mu W_\alpha + \bar R W^\sigma W_\sigma \right\} \nn \\
&&-16f^3 \left\{W^\mu W^\nu \bna_\mu W_\nu - W_\mu W^\mu \bna_\alpha W^\alpha \right\} +12 f^4 (W^\sigma W_\sigma )^2, \nn
\eea
\bea
R_{\mu\nu\lambda\alpha}R^{\lambda\alpha\mu\nu} &=& (\bar R_{\mu\nu\lambda\alpha} + Q_{\mu\nu\lambda\alpha})(\bar R^{\mu\nu\lambda\alpha} + Q^{\lambda\alpha\mu\nu}) \nn \\
&=& \bar R_{\mu\nu\lambda\alpha}\bar R^{\mu\nu\lambda\alpha}+2 Q_{\mu\nu\lambda\alpha}\bar R^{\mu\nu\lambda\alpha} + Q_{\mu\nu\lambda\alpha} Q^{\lambda\alpha\mu\nu} \nn \\
&=& (\bar R_{\mu\nu\lambda\alpha})^2 + 8 f \bar R^{\mu\alpha} \bna_\mu W_\alpha \\
&&+4f^2 \left\{ 2(\bna_\mu W_\nu)(\bna^\nu W^\mu)+(\bna_\mu W^\mu)^2 - 2 \bar R ^{\mu\alpha} W_\mu W_\alpha + \bar R W^\sigma W_\sigma \right\} \nn \\
&&-16f^3 \left\{W^\mu W^\nu \bna_\mu W_\nu - W_\mu W^\mu \bna_\alpha W^\alpha \right\} +12 f^4 (W^\sigma W_\sigma )^2 \nn \\
&=& (\bar R_{\mu\nu\lambda\alpha})^2 + 8 f \bar R^{\mu\alpha} \bna_\mu W_\alpha \\
&&+4f^2 \left\{ 2\bna_\mu (W_\nu\bna^\nu W^\mu)+(\bna_\mu W^\mu)^2 - 2  W^\nu \bna_\nu \bna_\mu W^\mu + \bar R W^\sigma W_\sigma \right\} \nn \\
&&-16f^3 \left\{W^\mu W^\nu \bna_\mu W_\nu - W_\mu W^\mu \bna_\alpha W^\alpha \right\} +12 f^4 (W^\sigma W_\sigma )^2, \nn
\eea
and the square of the Ricci tensors are
\bea
\left(R_{\mu\lambda}\right)^2 &=& (\bar R_{\mu\lambda} + Q_{\mu\lambda} )(\bar R^{\mu\lambda} + Q^{\mu\lambda} ) = \bar R_{\mu\lambda}\bar R^{\mu\lambda}+2 Q_{\mu\lambda}\bar R^{\mu\lambda} + Q_{\mu\lambda} Q^{\mu\lambda} \nn \\
&=& (\bar R_{\mu\lambda})^2 + 2f \left(2\bar R^{\mu\lambda} \bna_\mu W_\lambda + \bar R \bna_\alpha W^\alpha \right) \\ &&+f^2 \big\{ - 4 \bar R^{\mu\lambda}W_\mu W_\lambda + 4 \bar R W^\sigma W_\sigma +4 (\bna_\mu W_\lambda)^2 + 12 (\bna_{[\mu} W_{\lambda]})^2 \nn \\&& + 6(\bna_\alpha W^\alpha)^2 \big\}
-f^3 \left\{8W_\mu W_\lambda \bna^\mu W^\lambda - 20 (W_\sigma W^\sigma )(\bna_\alpha W^\alpha) \right\} \nn \\
&&+12 f^4 (W^\alpha W_\alpha)^2, \nn
\eea
\bea
(R_{\mu\lambda}&+&\breve R_{\mu\lambda})(R^{\lambda\mu}+\breve R^{\lambda\mu}) \nn \\
 &=& (2\bar R_{\mu\lambda} + Q_{\mu\lambda}+\breve Q_{\mu\lambda} )(2\bar R^{\mu\lambda} + Q^{\lambda\mu} +\breve Q^{\lambda\mu} )\\
&=& 4(\bar R_{\mu\lambda})^2 + 8f \left(2\bar R^{\mu\lambda} \bna_\mu W_\lambda + \bar R \bna_\alpha W^\alpha \right) \\ &&+16 f^2 \left\{ -  \bar R^{\mu\lambda}W_\mu W_\lambda +  \bar R W^\sigma W_\sigma + (\bna_\mu W_\lambda)(\bna^\lambda W^\mu)  + 2(\bna_\alpha W^\alpha)^2 \right\} \nn \\
&&-16f^3 \left\{2W_\mu W_\lambda \bna^\mu W^\lambda - 5 (W_\sigma W^\sigma )(\bna_\alpha W^\alpha) \right\} +48 f^4 (W^\alpha W_\alpha)^2, \nn
\eea
and the square of Ricci scalar is
\bea
R^2 &=& \left(\bar R + Q \right)^2 = \bar R^2 + 2 \bar R Q + Q^2 \nn \\
&=& \bar R^2 + 12 f \bar R \bna_\mu W^\mu + 12 f^2 \left( \bar R W^\mu W_\mu + 3 (\bna_\mu W^\mu)^2 \right) \\ && + 72f^3 W^\mu W_\mu \bna_\sigma W^\sigma + 36 f^4 (W^\mu W_\mu)^2. \nn
\eea

Now with these terms, let us think of Gauss-Bonnet like Lagrangian. This Lagrangian has the Weyl symmetry and reduces the topological invariant in Riemann space-time.
\bea
\mathcal{L}_{GB} &=& - \dfrac{1}{4}\left\{(R_{\mu\nu\lambda\alpha})^2 - 4 (R_{\mu\lambda})^2+R^2 \right\} \nn \\
&=& - \dfrac{1}{4}\Big\{(\bar R_{\mu\nu\lambda\alpha})^2 - 4 (\bar R_{\mu\lambda})^2+\bar R^2 -f\left( 8\bar R^{\mu\nu}\bna_\mu W_\nu - 4\bar R \bna_\mu W^\mu \right)  \nn \\
&& +f^2 \left( 8\bar R^{\mu\nu} W_\mu W_\nu - 8 (\bna_\mu W_\nu)^2 -32 (\bna_{[\mu}W_{\nu]})^2 +16(\bna_\mu W^\mu)^2 \right) \nn \\
&& +f^3 \left(16W^\mu W^\nu \bna_\mu W_\nu + 8 W^\mu W_\mu \bna_\alpha W^\alpha \right) \Big\}
\eea
Using $(\bna_\mu W_\nu )^2 = 2(\bna_{[\mu}W_{\nu]})^2 +R_{\mu\nu} W^\mu W^\nu + \bna_\mu (W^\nu \bna_\nu W^\mu ) -W^\nu \bna_\nu \bna_\mu W^\mu$, we can express the square terms like following
\bea
\mathcal{L}_{GB}^{(2)} &=& - \dfrac{1}{4} f^2 \Big\{ -48(\bna_{[\mu}W_{\nu]})^2 + 16(\bna_\mu W^\mu)^2 - 8\bna_\mu (W^\nu \bna_\nu W^\mu)+ 8 W^\nu \bna_\nu \bna_\mu W^\mu \Big\} \nn \\
&=&- \dfrac{1}{4} f^2 \Big\{ -12F_{\mu\nu}F^{\mu\nu} + 16(\bna_\mu W^\mu)^2 - 8 \bna_\mu (W^\nu \bna_\nu W^\mu)+ 8 W^\nu \bna_\nu \bna_\mu W^\mu \Big\} \nn
\eea
where $F_{\mu\nu}\equiv \bna_\mu W_\nu - \bna_\nu W_\mu $. Unfortunately this Lagrangian has the negative kinetic energy terms of the Weyl vector fields. So it is not interesting. Of course this Lagrangian is not the topological invariant in Weyl geometry. In general the Weyl-Cartan geometry has the topological invariant quantity slightly different from the Gauss-Bonnet identity\cite{BF}. It has the form like following
\bea
I_{BF} &\equiv& R^2 - (R_{\mu\nu}+ \breve R_{\mu\nu})(R^{\nu\mu}+ \breve R^{\nu\mu})+R_{\mu\nu\alpha\beta} R^{\alpha\beta\mu\nu}.
\eea
So the topological invariant Lagrangian is
\bea
\mathcal{L}_{BF} &=& - \dfrac{1}{4}\left\{(R_{\mu\nu\lambda\alpha})^2 - (R_{\mu\nu}+ \breve R_{\mu\nu})(R^{\nu\mu}+ \breve R^{\nu\mu})+R^2 \right\} \nn \\
&=& - \dfrac{1}{4}\Big\{(\bar R_{\mu\nu\lambda\alpha})^2 - 4 (\bar R_{\mu\lambda})^2+\bar R^2 -f\left( 8\bar R^{\mu\nu}\bna_\mu W_\nu - 4\bar R \bna_\mu W^\mu \right)  \nn \\
&& +f^2 \left( 8\bar R^{\mu\nu} W_\mu W_\nu - 8 (\bna_\mu W_\nu)(\bna^\nu W^\mu) +8(\bna_\mu W^\mu)^2 \right) \nn \\
&& +f^3 \left(16W^\mu W^\nu \bna_\mu W_\nu + 8 W^\mu W_\mu \bna_\alpha W^\alpha \right) \Big\} \\
&=& - \dfrac{1}{4}\Big\{(\bar R_{\mu\nu\lambda\alpha})^2 - 4 (\bar R_{\mu\lambda})^2+\bar R^2 -f\left( 8\bar R^{\mu\nu}\bna_\mu W_\nu - 4\bar R \bna_\mu W^\mu \right)  \nn \\
&& +f^2 \left( 8 W_\nu\bna^\nu \bna_\mu W^\mu - 8 \bna_\mu (W_\nu\bna^\nu W^\mu) +8(\bna_\mu W^\mu)^2 \right) \nn \\
&& +f^3 \left(16W^\mu W^\nu \bna_\mu W_\nu + 8 W^\mu W_\mu \bna_\alpha W^\alpha \right) \Big\} \nn
\eea
But if we apply the gauge fixing condition $\bna_\mu W^\mu = 0$ to this Lagrangian, the square terms has no dynamics. So this is also out of interesting.
Now let us think of another the Yang-Mills type Lagrangian.
\bea
\mathcal{L}&=& - \dfrac{1}{4} (R_{\mu\nu\lambda\alpha})^2 \nn \\
&=& - \dfrac{1}{4} \Big\{
(\bar R_{\mu\nu\lambda\alpha})^2 + 8 f \bar R^{\mu\alpha} \bna_\mu W_\alpha \\
&&+4f^2 \big( 8(\bna_{[\mu} W_{\nu]})^2 + (\bna_\mu W^\mu)^2  + \bar R W^\sigma W_\sigma
 +2 \bna_\mu (W^\nu \bna_\nu W^\mu )  \nn \\ &&- 2W^\nu \bna_\nu \bna_\mu W^\mu\big)
-16f^3 \left(W^\mu W^\nu \bna_\mu W_\nu - W_\mu W^\mu \bna_\alpha W^\alpha \right) +12 f^4 (W^\sigma W_\sigma )^2
\Big\} \nn
\eea
And this Lagrangian has the positive kinetic energy terms. If we omit the total divergence terms and $\bna_\mu W^\mu$ terms which vanish under the gauge fixing condition $\bna_\mu W^\mu = 0$. Then the Lagrangian becomes
\bea
\mathcal{L'}&=& - \dfrac{1}{4} \Big\{
(\bar R_{\mu\nu\lambda\alpha})^2 + 8 f \bar R^{\mu\alpha} \bna_\mu W_\alpha +4f^2 \left( 4(\bna_{\mu} W_{\nu})^2 - 4 \bar R_{\mu\nu}W^\mu W^\nu  + \bar R W^\sigma W_\sigma
  \right) \nn \\
&&-16f^3 W^\mu W^\nu \bna_\mu W_\nu  +12 f^4 (W^\sigma W_\sigma )^2
\Big\} \nn \\
&=& - \dfrac{1}{4}(\bar R_{\mu\nu\lambda\alpha})^2 -2 f \bar R^{\mu\alpha} \bna_\mu W_\alpha -4f^2 \left( (\bna_{\mu} W_{\nu})^2 -  \bar R_{\mu\nu}W^\mu W^\nu  + \dfrac{1}{4} \bar R W^\sigma W_\sigma
  \right) \nn \\
&&+4f^3 W^\mu W^\nu \bna_\mu W_\nu  -3 f^4 (W^\sigma W_\sigma )^2. \label{torsionfreeYM}
\eea
So in torsion free case, it is desirable to use the Yang-Mills type Lagrangian rather than the Gauss-Bonnet type.

%===================================================================

\section{One-loop effective actions}
Now let us calculate the one-loop effective actions. Exact calculation of the effective action for an arbitrary curvature is very hard to solve, so we will consider the constant curvature cases with Riemann normal coordinates\cite{normal coord}.

First let us consider the Lagrangian (\ref{torsionfreeYM}) which is the torsion free Yang-Mills type Lagrangian.
To calculate the effective action we should split the Weyl vector field into the "classical" part $W^0_\mu$ and the quantum fluctuating part $W^q_\mu$. But the vacuum Weyl field condensate $<0|Q_{\mu\nu\lambda}^{~~~\kappa}|0>$ should be the form of a gauge covariant additive combination $\bar R_{\mu\nu\lambda}^{~~~\kappa}+<\!Q_{\mu\nu\lambda}^{~~~\kappa}\!>$. So, to find the functional dependence of the
effective potential $V_{eff} (\bar R+<\!Q\!>)$
on $<\!Q\!>$
we calculate first the effective potential $V_{eff} (\bar R)$
by setting $\tilde W^0_{\mu}=0$, i.e., $W_{\mu}=W^q_{\mu}$.
Then, after completing the calculation
we will restore the dependence on Weyl field condensate
$<\!Q\!>$ by simple adding this term
to $\bar R$ in the final expression for
$V_{eff} (\bar R)$\cite{pak}.

With the Weyl gauge fixing condition
$\bna_\mu W^\mu=0$ and $\delta(\bna_\mu W^\mu) = -\frac{1}{f}(\bna^\mu -4 f W^\mu)\pro_\mu \Lambda$, one can find the
gauge fixing term ${\cal L}_{GF}$ and Faddeev-Popov ghost term
${\cal L}_{FP}$
\bea
{\cal L}_{GF} &=& - \dfrac{1}{2\xi} (\bna_\mu W^\mu)^2, \\
{\cal L}_{FP} &=& \bar {c} \bna^\mu (\bna_\mu {c}) -4 f \bar {c} W^\mu(\bna_\mu {c}).
\eea
So the total quadratic Lagrangian becomes
\bea
\mathcal{L'}
&=& - \dfrac{1}{4}(\bar R_{\mu\nu\lambda\alpha})^2 -4f^2 \left( (\bna_{\mu} W_{\nu})^2 -  \bar R_{\mu\nu}W^\mu W^\nu  + \dfrac{1}{4} \bar R W^\sigma W_\sigma
  \right) \nn \\
&&- \dfrac{1}{2\xi} (\bna_\mu W^\mu)^2 + \bar {c} \bna^\mu(\bna_\mu {c})-4 f \bar {c} W^\mu(\bna_\mu {c})
\eea

Now let us think of the Riemann normal coordinates in constant curvature. The curvature tensor becomes
\be
\bar R_{\mu\nu\alpha\beta} =  \dfrac{1}{12}\bar R (g_{\mu\alpha}g_{\nu\beta} - g_{\mu\beta}g_{\nu\alpha}).\label{constcurvature}
\ee
Let the Riemann normal coordinates and metric be $x^\mu$ and $g_{\mu\nu}$ for the point $P$ respectively.
And let the metric at origin be $\eta_{\mu\nu}$. Then  in the case of $\bna_\alpha \bar R_{\mu\nu\lambda\kappa} =0$ the metric at the point $P$ may be expanded by \cite{normal coord}
\bea
g_{\mu\nu}(x)&=&\eta_{\mu\nu} + \dfrac{1}{2} \sum _{k=1}^{\infty} \dfrac{2^{2k+2}}{(2k+2)!}f_{\mu\sigma_1} f_{\sigma_2}^{\sigma_1}\cdots f^{\sigma_{k-1}}_{\nu}   \nn \\
&=&\eta_{\mu\nu} + \frac{1}{3} \bar R_{\mu\alpha\nu\beta}x^\alpha x^\beta + \frac{2}{45}\bar R_{\alpha\mu\beta\sigma}\bar R_{\gamma\nu\delta}^{~~~\sigma}x^\alpha x^\beta x^\gamma x^\delta + \cdots \label{metric}
\eea
where $f^{\sigma}_{\mu}=\bar R_{\alpha\mu\beta}^{~~~{\sigma}} x^\alpha x^\beta$
and $f_{\sigma\nu}=\bar R_{\alpha\sigma\beta\nu} x^\alpha x^\beta$.

Now we want to put eq.(\ref{constcurvature}) into eq.(\ref{metric}). Since
\bea
f_{\mu}^{\sigma}&=&\bar R_{\alpha\mu\beta}^{~~~{\sigma}} x^\alpha x^\beta= \frac{1}{12} \bar R(g_\mu^\sigma x_\alpha x^\alpha - x_\mu x^\sigma) = \frac{1}{12} \bar R(g_\mu^\sigma x^2 - x_\mu x^\sigma) \nn\\
f_{\mu\sigma}&=&\bar R_{\alpha\sigma\beta\nu} x^\alpha x^\beta =\frac{1}{12} \bar R(g_{\mu\sigma} x^2 - x_\mu x_\sigma)=g_{\sigma\alpha}f_\mu^\alpha, \nn
\eea
then $f_{\mu}^{\sigma} x^\mu = \frac{1}{12} \bar R(x^\sigma x^2 - x^2 x^\sigma)=0$. So $x_\mu = g_{\mu\nu} x^{\nu} = \eta_{\mu\nu} x^{\nu}$ and $x^2 = g_{\mu\nu} x^\mu x^{\nu} = \eta_{\mu\nu} x^\mu x^{\nu}$. And
\bea
f_{\mu\sigma}f_{\nu}^{\sigma}&=& \Big(\dfrac{\bar R}{12} \Big)^2 (g_{\mu\sigma}x^2 - x_\mu x_\sigma)(g_{\nu}^{\sigma}x^2 - x_\nu x^\sigma) \nn \\
&=& \Big(\dfrac{\bar R}{12} \Big)^2 x^2 (g_{\mu\nu}x^2 - x_\mu x_\nu) = \dfrac{\bar R}{12} x^2 f_{\mu\nu} \nn \\
f_{\mu\sigma_1}f_{\sigma_2}^{\sigma_1}f_{\nu}^{\sigma_2} &=& \dfrac{\bar R}{12} x^2 f_{\mu\sigma_2}f_{\nu}^{\sigma_2} = \dfrac{\bar R}{12} x^2 \cdot \dfrac{\bar R}{12} x^2 f_{\mu\nu} = \Big(\dfrac{\bar R}{12} x^2 \Big)^2 f_{\mu\nu} \nn \\
&\vdots& \nn \\
f_{\mu\sigma_1} f_{\sigma_2}^{\sigma_1}\cdots f^{\sigma_{k-1}}_{\nu} &=& \Big(\dfrac{\bar R}{12} x^2 \Big)^{k-1} f_{\mu\nu} \nn
\eea
And putting these results into eq.(\ref{metric}), we get
\bea
g_{\mu\nu}(x)&=&\eta_{\mu\nu} + \dfrac{1}{2} \sum _{k=1}^{\infty} \dfrac{2^{2k+2}}{(2k+2)!}f_{\mu\sigma_1} f_{\sigma_2}^{\sigma_1}\cdots f^{\sigma_{k-1}}_{\nu}   \nn \\
&=&\eta_{\mu\nu} +  \dfrac{1}{2} \sum _{k=1}^{\infty} \dfrac{2^{2k+2}}{(2k+2)!}\Big(\dfrac{\bar R}{12} x^2 \Big)^{k-1} f_{\mu\nu}  \\
&=&\eta_{\mu\nu} + \dfrac{72\cosh\sqrt{\frac{\bar R x^2}{3}}-12\bar R x^2 -72}{\bar R^2 x^4}  f_{\mu\nu} \\
&=&\eta_{\mu\nu} + \dfrac{72\cosh\sqrt{\frac{\bar R x^2}{3}}-12\bar R x^2 -72}{\bar R^2 x^4}\cdot \frac{1}{12} \bar R(g_{\mu\nu} x^2 - x_\mu x_\nu) \\
&=&\eta_{\mu\nu} + \dfrac{6\cosh\sqrt{\frac{\bar R x^2}{3}}-\bar R x^2 -6}{\bar R x^4}(g_{\mu\nu} x^2 - x_\mu x_\nu)\label{metric2}
\eea
where we have used $\frac{1}{2} \sum _{k=1}^{\infty} \frac{2^{2k+2}}{(2k+2)!}\Big(\frac{\bar R}{12} x^2 \Big)^{k-1} = \frac{72\cosh\sqrt{\frac{\bar R x^2}{3}}-12\bar R x^2 -72}{\bar R^2 x^4}$.

Now we can solve the eq.(\ref{metric2}) for $g_{\mu\nu}$ and the final expression of the metric in terms of Riemann normal coordinate with eq.(\ref{constcurvature}) becomes
\bea
g_{\mu\nu}(x) = \dfrac{\bar R x^2 }{2\bar R x^2 - 12 \sinh^2 \sqrt{\frac{\bar R x^2}{12}}}\left(\eta_{\mu\nu} - \dfrac{12 \sinh^2 \sqrt{\frac{\bar R x^2}{12}}-\bar R x^2}{\bar R x^4}x_\mu x_\nu  \right).
\eea
And the inverse of this metric is
\bea
g^{\mu\nu}(x) = \eta^{\mu\nu} + \dfrac{12 \sinh^2 \sqrt{\frac{\bar R x^2}{12}}-\bar R x^2}{\bar R x^4}(x^\mu x^\nu -\eta^{\mu\nu} x^2).
\eea
The determinant of $g_{\mu\nu}$ is $\det(g_{\mu\nu})=\Big(\frac{\bar R x^2 }{2\bar R x^2 - 12 \sinh^2 \sqrt{\frac{\bar R x^2}{12}}}\Big)^3\det(\eta_{\mu\nu}) $.

Next calculating the Christoffel symbols, they are
\bea
\bar \Gamma_{\mu\nu\lambda}(x) &=& \frac{1}{2} (\partial_\mu g_{\nu\lambda} +\pro_\nu g_{\mu\lambda}-\pro_\lambda g_{\mu\nu}) \nn \\
&=&\left(\dfrac{1}{x^2}-\dfrac{\bar R}{\bar R x^2 - 6\sinh^2 \sqrt{\frac{\bar R x^2}{12}}}+\dfrac{2\bar R^2 x^2 - \sqrt{3\bar R^3 x^2}\sinh\sqrt{\frac{\bar R x^2}{3}}}{4 \left(\bar R x^2 - 6\sinh^2 \sqrt{\frac{\bar R x^2}{12}}\right)^2}\right)\eta_{\mu\nu}x_\lambda \nn \\
&&+\left(-\dfrac{1}{x^4}+\dfrac{2\bar R^2 \sqrt{x^2} - \sqrt{3\bar R^3 }\sinh\sqrt{\frac{\bar R x^2}{3}}}{4 \sqrt{x^2} \left(\bar R x^2 - 6\sinh^2 \sqrt{\frac{\bar R x^2}{12}}\right)^2}\right)x_{\mu}x_{\nu}x_\lambda  \\
&&+\left(\dfrac{\bar R \left( \sqrt{3\bar R x^2 }\sinh\sqrt{\frac{\bar R x^2}{3}}-12\sinh^2 \sqrt{\frac{\bar R x^2}{12}}\right)}{4 \sqrt{x^2} \left(\bar R x^2 - 6\sinh^2 \sqrt{\frac{\bar R x^2}{12}}\right)^2}\right)(x_{\mu}\eta_{\nu\lambda} + x_\nu \eta_{\mu\lambda}) \nn
\eea
and
\bea
\bar \Gamma_{\mu\nu}^{~~\kappa}(x)&=&g^{\kappa\lambda}\bar \Gamma_{\mu\nu\lambda}(x) \nn \\
&=&\left(\dfrac{1}{x^2}-\dfrac{\bar R}{\bar R x^2 - 6\sinh^2 \sqrt{\frac{\bar R x^2}{12}}}+\dfrac{2\bar R^2 x^2 - \sqrt{3\bar R^3 x^2}\sinh\sqrt{\frac{\bar R x^2}{3}}}{4 \left(\bar R x^2 - 6\sinh^2 \sqrt{\frac{\bar R x^2}{12}}\right)^2}\right)\eta_{\mu\nu}x^\kappa \nn \\
&&+\left(\dfrac{\bar R \left( \sqrt{3\bar R x^2 }\cosh\sqrt{\frac{\bar R x^2}{12}}-6\sinh \sqrt{\frac{\bar R x^2}{12}}\right)}{2 {x^4} \left(\bar R x^2 - 6\sinh^2 \sqrt{\frac{\bar R x^2}{12}}\right)^2}\right)(x_{\mu}\delta_{\nu}^{\kappa} + x_\nu \delta_{\mu}^{\kappa}) \nn \\
&&~~~~\cdot\left\{(9\bar R  + \bar R^2 x^2 + x^4) \sinh\sqrt{\frac{\bar R x^2}{12}}-3\bar R \sinh\sqrt{\frac{\bar R x^2}{4}} \right\} \nn  \\
&&-\dfrac{1}{4 {x^4} \left(\bar R x^2 - 6\sinh^2 \sqrt{\frac{\bar R x^2}{12}}\right)^2}\Bigg \{
2(9+\bar R^2 x^2 )^2  \\
&&~~~~ - 36 (6+ \bar R x^2 ) \cosh \sqrt{\frac{\bar R x^2}{3}} + 54\cosh\sqrt{\frac{4\bar R x^2}{3}} \nn \\
&& ~~~~ + 3\sqrt{3 \bar R x^2} \Big( \bar R x^2 - 8\sinh \sqrt{\frac{\bar R x^2}{12}}\Big) \sinh \sqrt{\frac{\bar R x^2}{3}}\Bigg\}x_{\mu}x_{\nu}x^\kappa \nn
\eea

Now with above equipments, let us compute the effective action. With constant curvature the effective action can be written in the form
\bea
\exp \big(i\Gamma_{eff} \big) &=&
\int
{\cal D} { W}_\mu {\cal D} { c} {\cal D} \bar { c}  \\
&&\exp \Big{\lbrace} i\int d^4x\sqrt {-g}  \Big[ - \dfrac{1}{4}(\bar R_{\mu\nu\lambda\alpha})^2 -4f^2 (\bna_{\mu} W_{\nu})^2  + \bar {c} \bna^\mu(\bna_\mu {c}) \Big] \Big{\rbrace} .\nn
\eea
In flat space-time, to calculate functional integration we can use following
\be
\int {\cal D \phi} e^{-\int d^4x d^4y \phi(x)\cdot A(x,y)\cdot \phi(y)} = (\det A)^{-1/2}
\ee
for any real {\it symmetric}, positive, non-singular matrix.
But in curved space-time it is not trivial to make the symmetric matrix. For the scalar fields,
\bea
&&\int d^4 x \sqrt{-g(x)}\pro_\mu \phi(x) \pro_\nu \phi(x) g^{\mu\nu}(x)\nn \\
&&=\int d^4 x \sqrt{-g(x)}\int d^4 y \sqrt{-g(y)}\nn \\
&&~~\cdot \pro^x_\mu \phi(x) \pro^y_\nu \phi(y) g^{\mu\nu}(x,y)\delta^4(x,y)(-g(x))^{-\frac{1}{4}}(-g(y))^{-\frac{1}{4}}\nn \\
&&=-\int d^4 x \int d^4 y \phi(x) \pro^x_\mu \big\{ (-g(x))^{\frac{1}{4}}(-g(y))^{\frac{1}{4}}\pro^y_\nu \phi(y) g^{\mu\nu}(x,y)\delta^4(x,y) \big\} \nn \\
&&=\int d^4 x \int d^4 y \phi(x)\phi(y) \pro^x_\mu \pro^y_\nu\big\{ (-g(x))^{\frac{1}{4}}(-g(y))^{\frac{1}{4}}  g^{\mu\nu}(x,y)\delta^4(x,y) \big\} \nn \\
&&=\int d^4 x \int d^4 y \sqrt{-g(x)}\sqrt{-g(y)}\phi(x)\phi(y) \\
  && ~~\cdot \frac{1}{\sqrt{-g(x)}\sqrt{-g(y)}}\pro^x_\mu \pro^y_\nu\big\{ (-g(x))^{\frac{1}{4}}(-g(y))^{\frac{1}{4}}  g^{\mu\nu}(x,y)\delta^4(x,y) \big\} \nn
\eea
where in $g^{\mu\nu}(x,y)$, the variables of $g^{\mu\nu}(x)$ are changed symmetrically by the rule $x^\mu x^\nu \rightarrow \frac{x^\mu y^\nu+y^\mu x^\nu}{2}$. So we have got the symmetric matrix which is defined by
\bea
A_{scalar} (x,y) \equiv  \frac{1}{\sqrt{-g(x)}\sqrt{-g(y)}}\pro^x_\mu \pro^y_\nu\big\{ (-g(x))^{\frac{1}{4}}(-g(y))^{\frac{1}{4}}  g^{\mu\nu}(x,y)\delta^4(x,y) \big\}
\eea
After some integration by parts of Dirac delta function with Riemann normal coordinates and neglecting the total divergence, we get following in the limit of $x^\mu, y^\mu\rightarrow 0$
\bea
A_{scalar} (x,y)&\simeq& ( \pro^x_\mu \pro_y^\mu + \frac{1}{6}\bar R)\delta^4(x,y) \nn\\
&=& \int \frac{d^4 p}{(2\pi)^4} (\pro^x_\mu \pro_y^\mu + \frac{1}{6}\bar R)e^{-i p \cdot (x-y)} \\
&=&\int \frac{d^4 p}{(2\pi)^4} (p_\mu p^\mu + \frac{1}{6}\bar R)e^{-i p \cdot (x-y)}\nn
\eea
Therefore
\bea
{\rm Tr} \ln A_{scalar}(x,y) &=& \int d^4x d^4y \delta^4(x-y)\ln A_{scalar}(x,y) \nn \\
&\simeq& \int d^4x \int \frac{d^4 p}{(2\pi)^4} \ln (p^2 + \frac{1}{6}\bar R)
\eea
For the vector fields,
\bea
&&\int d^4 x \sqrt{-g(x)} g^{\mu\nu}(x)g_{\alpha\beta}(x)\bna_\mu W^\alpha(x) \bna_\nu W^\beta(x) \nn \\
&&~~~~~=\int d^4 x \sqrt{-g(x)}\int d^4 y \sqrt{-g(y)}\nn \\
&& ~~~~~~~~~~~g^{\mu\nu}(x,y)g_{\alpha\beta}(x,y)
\bna^x_\mu W^\alpha(x) \bna^y_\nu W^\beta(y) \frac{\delta^4(x,y)}{(-g(x))^{\frac{1}{4}}(-g(y))^{\frac{1}{4}}} \nn \\
&&~~~~~\simeq \int d^4 x \int d^4 y  W^\alpha(x) W^\beta(y)\eta_{\alpha\beta}(\pro^x_\mu \pro_y^\mu + \frac{1}{12} \bar R)\delta^4(x,y)
\eea
Therefore,
\bea
A_{\alpha\beta}(x,y)&\simeq& \eta_{\alpha\beta}(\pro^x_\mu \pro_y^\mu + \frac{1}{12} \bar R)\delta^4(x,y)\nn \\
&=&\int \frac{d^4 p}{(2\pi)^4} \eta_{\alpha\beta}(p^2 + \frac{1}{12}\bar R)e^{-i p \cdot (x-y)}.
\eea
and
\bea
{\rm Tr} \ln A_{\alpha\beta}(x,y) &=& \int d^4x d^4y \delta^4(x-y){\rm tr}\ln A_{\alpha\beta}(x,y) \nn \\
&\simeq& 4\int d^4x \int \frac{d^4 p}{(2\pi)^4} \ln (p^2 + \frac{1}{12}\bar R)
\eea

Now the effective action is given by
\bea
 \Gamma_{eff}=\frac{i}{2} {\rm Tr} \ln A_{\alpha\beta}(x,y) - i {\rm Tr} \ln A_{scalar} (x,y)
\eea
Neglecting overall constant we have one-loop effective potential
\bea
\Gamma_{(1)} = \int d^4 x \int \frac{d^4 p}{(2\pi)^4}\{2 i  \ln (1+ \frac{\bar R}{12 p^2})-i  \ln (1+ \frac{\bar R}{6 p^2}) \} \equiv i \int d^4 x \mathcal{L}^{(1)}_{eff}
\eea
Then,
\bea
\mathcal{L}^{(1)}_{eff} = 2 \int \frac {d^4 p}{(2\pi)^4}\ln (1+\frac{\bar R}{12 p^2}) -\int \frac {d^4 p}{(2\pi)^4}\ln (1+\frac{\bar R}{6 p^2})
\eea
Since
\bea
\int \frac {d^4 p}{(2\pi)^4}\ln (1+\frac{a}{ p^2}) &=& \frac {2\pi^2}{(2\pi)^4}\int_0^\Lambda dp ~p^3 \ln (1+\frac{a}{ p^2}) \nn \\
&=&2\cdot \frac{\Lambda^2}{32\pi^2} a + \frac{a^2}{32\pi^2}(\ln \frac{a}{\Lambda^2} -\frac{1}{2})+(\cdots)
\eea
where the last term vanishes when $\Lambda\rightarrow \infty$,
\bea
\mathcal{L}^{(1)}_{eff} &\simeq& 2\cdot 2\cdot \frac{\Lambda^2}{32\pi^2} (\frac{\bar R}{12}) + 2 (\frac{\bar R}{12})^2\frac{1}{32\pi^2}(\ln \frac{1}{\Lambda^2} \cdot\frac{\bar R}{12} -\frac{1}{2}) \nn \\
&&-2\cdot \frac{\Lambda^2}{32\pi^2} (\frac{\bar R}{6}) - \frac{1}{32\pi^2}(\frac{\bar R}{6})^2(\ln \frac{1}{\Lambda^2}\cdot \frac{\bar R}{6} -\frac{1}{2})\nn \\
&=& \frac{1+2\ln 3}{4608\pi^2}\bar R^2 - \frac {1}{2304\pi^2} \bar R^2
\ln \frac{\bar R}{\Lambda^2} \\
&=& \frac{1+2\ln 3}{768\pi^2}\bar R_{\mu\nu\alpha\beta}^2 - \frac {1}{384\pi^2} \bar R_{\mu\nu\alpha\beta}^2
\ln \frac{\sqrt{6\bar R_{\mu\nu\alpha\beta}^2}}{\Lambda^2} \nn
\eea

Now by shifting $\bar R_{\mu\nu\alpha\beta}\rightarrow \bar R_{\mu\nu\alpha\beta}+ <\!Q_{\mu\nu\alpha\beta}\!>$ we get,
\bea
\mathcal{L}^{(1)}_{eff} &\simeq& \dfrac{1+2\ln 3}{768\pi^2}(\bar R_{\mu\nu\alpha\beta}+ <\!Q_{\mu\nu\alpha\beta}\!>)^2 \nn \\ &&- \dfrac {1}{384\pi^2} (\bar R_{\mu\nu\alpha\beta}+ <\!Q_{\mu\nu\alpha\beta}\!>)^2
\ln \frac{\sqrt{6(\bar R_{\mu\nu\alpha\beta}+ <\!Q_{\mu\nu\alpha\beta}\!>)^2}}{\Lambda^2} \\
&=& - \dfrac {1}{384\pi^2} (\bar R_{\mu\nu\alpha\beta}+ <\!Q_{\mu\nu\alpha\beta}\!>)^2 \left\{
\ln \dfrac{\sqrt{(\bar R_{\mu\nu\alpha\beta}+ <\!Q_{\mu\nu\alpha\beta}\!>)^2}}{\Lambda^2} +\ln \dfrac{\sqrt6}{3} - \dfrac{1}{2}\right\} \nn
\eea
 Note that this is not exact but only approximation in the constant curvature background.
With normalization condition $\frac{\partial^2 \mathcal{V}_{eff}}{(\partial R_{\mu\nu\alpha\beta})^2}\Big|_{R_{\mu\nu\alpha\beta}=\Lambda} =\frac{1}{2}$ the renormalized effective potential is now
\bea
\mathcal{V}_{eff} &\simeq& \dfrac{1}{4}(\bar R_{\mu\nu\alpha\beta}+ <\!Q_{\mu\nu\alpha\beta}\!>)^2 \\
&&+\dfrac {1}{384\pi^2} (\bar R_{\mu\nu\alpha\beta}+ <\!Q_{\mu\nu\alpha\beta}\!>)^2 \left\{
\ln \dfrac{\sqrt{(\bar R_{\mu\nu\alpha\beta}+ <\!Q_{\mu\nu\alpha\beta}\!>)^2}}{\Lambda^2}  - \dfrac{3}{2}\right\} \nn
\eea
This potential has the minimum value $\mathcal{V}_{min}$ when $\bar R_{\mu\nu\alpha\beta}=0$ and $<\!Q_{\mu\nu\alpha\beta}\!>\neq0$.
\bea
&&\mathcal{V}_{min} = - \dfrac{1}{768 \pi^2} <\!Q_{\mu\nu\alpha\beta}\!>^2, \\
&&<\!Q_{\mu\nu\alpha\beta}\!>= e^{1-96\pi^2}\Lambda^2
\eea
Expanding the original classical Lagrangian around the new vacuum we obtain
\bea
\mathcal{L}_{eff} &\simeq& -\dfrac{1}{4}(\bar R_{\mu\nu\alpha\beta}+ <\!Q_{\mu\nu\alpha\beta}\!>)^2
 =-\dfrac{1}{4} \bar R_{\mu\nu\alpha\beta}^2 - \dfrac{1}{2} \bar R M^2
-\dfrac{3}{2} M^4.\label{LQeff}
\eea
where we put $<Q_{\mu\nu\alpha\beta}> = \dfrac{1}{2} M^2 (g_{\mu\alpha} g_{\nu\beta}-g_{\mu\beta}g_{\nu\alpha})$ \cite{pak}. So we can get the Einstein-Hilbert type terms in the effective Lagrangian
(in units $\hbar = c=1$)
\bea
{\cal L}_{EHeff}=- \difrac{1}{4} \bar R_{\mu\nu\alpha\beta}^2 - \difrac{1}{16 \pi G} (\bar R
+2 \lambda) .
\eea
Thus we have similar result with the Weyl fields as with the torsion case in \cite{pak}. This means that the Weyl fields can be important player of quantum gravity like the torsion fields. And in low energy limit the metric gets the dynamics and Einstein-Hilbert term becomes dominant. This Einstein-Hilbert term contains only metric field without torsion and Weyl vector fields. So this describes the conventional general relativity.
However if we introduce the scalar fields with non-minimal coupling term $\xi R \phi^2$ and integrate out with respect to the scalars then we get the full Einstein-Hilbert term which contains not only metric but also torsion and Weyl vector fields.
The Lagrangian of scalar field together with all permissible non-minimal coupling is given by \cite{odints}
\bea
\mathcal{L}_{scalar} &=& -\frac{1}{2} g^{\alpha\beta}\pro_\alpha \phi \pro_\beta \phi -\frac{1}{2} m^2 \phi^2 - \frac{\lambda}{4!}\phi^4 \\
&&+\frac{1}{2}(\xi_1 R +\xi_2 \bna_\mu K^\mu +\xi_3 K_\mu K^\mu + \xi_4 S_\mu S^\mu + \xi_5 M_{\mu\nu\lambda}M^{\mu\nu\lambda})\phi^2  \nn
\eea
where $K_\mu = K^{\sigma}_{~\sigma \mu}$, $S^\mu = \tilde \epsilon ^{\mu\nu\lambda\kappa} K_{\nu\lambda\kappa} $, $M^{\sigma}_{~\sigma \mu}=0$ and $\tilde \epsilon ^{\mu\nu\lambda\kappa} M_{\nu\lambda\kappa}=0$.

But for the conformal invariant Lagrangian we set $\xi_1 =\frac{1}{6}$, $m=\xi_i =0$ ($i=2,3,4,5$).
So the conformal invariant scalar Lagrangian becomes
\bea
\mathcal{L}_{scalar} = -\frac{1}{2} g^{\alpha\beta}\pro_\alpha \phi \pro_\beta \phi +\frac{1}{12} R \phi^2 - \frac{\lambda}{4!}\phi^4
\eea

In next section we will investigate the relation between the torsion and Weyl fields with the full Einstein-Hilbert action using Palatini formalism.

%-----------------------------------------------------
%\newpage
\section{General $R^2$-type Weyl-Cartan Gravity with the Palatini connections}\label{generalweyl}
\subsection{The Palatini Connections}
Palatini's approach is the first order formalism treating the
metric and connection as independent degrees of freedom and
varying separately with respect to them. With this method we can
naturally derive Weyl gravity from Einstein-Hilbert action.

Now we consider the low energy case, so let us start from Einstein-Hilbert Lagrangian density
\begin{equation}\label{eq:EH lagrangian}
    \mathcal{L} = \sqrt{-g}R
\end{equation}
and take variation with respect to the connections.

\bea
\dfrac{1}{\sqrt{-g}}\dfrac{\delta \mathcal{L}}{\delta \Gamma_{\mu \nu}^{~~\lambda}}
&=&\pro_\lambda g^{\mu\nu} - \delta_\lambda^\nu \pro_\rho g^{\mu\rho}+\dfrac{1}{2}g^{\mu\nu}\pro_\lambda g_{\sigma\kappa}\cdot g^{\sigma\kappa} \nn \\
&&-\dfrac{1}{2}\delta^{\nu}_\lambda\pro^\mu g_{\sigma\kappa}\cdot g^{\sigma\kappa} -g^{\mu\nu}\Gamma_{\sigma\lambda}^{~~\sigma}
+ g^{\rho\mu}\Gamma_{\rho\lambda}^{~~\nu}-\delta_\lambda^\nu g^{\rho\sigma}\Gamma_{\sigma\rho}^{~~\mu}+g^{\nu\rho}\Gamma_{\lambda\rho}^{~~\mu}, \nn \\
&=&0 \nn
\eea
and using the following identity
\bea
\nabla_\lambda g^{\mu\nu} &=& \pro_\lambda g^{\mu\nu} + \Gamma_{\lambda\kappa}^{~~\mu} g^{\kappa\nu} + \Gamma_{\lambda\kappa}^{~~\nu} g^{\mu\kappa} \nn \\
\nabla_\rho g^{\mu\rho} &=& \pro_\rho g^{\mu\rho} + \Gamma_{\rho\kappa}^{~~\mu} g^{\kappa\rho} + \Gamma_{\rho\kappa}^{~~\rho} g^{\mu\kappa} \nn \\
\nabla_\lambda g_{\sigma\kappa} &=& \pro_\lambda g_{\sigma\kappa} - \Gamma_{\lambda\sigma}^{~~\alpha} g_{\alpha\kappa} - \Gamma_{\lambda\kappa}^{~~\alpha} g_{\sigma\alpha} \nn \\
\nabla^\mu g_{\sigma\kappa} &=& g^{\mu\beta}(\pro_\beta g_{\sigma\kappa} - \Gamma_{\beta\sigma}^{~~\alpha} g_{\alpha\kappa} - \Gamma_{\beta\kappa}^{~~\alpha} g_{\sigma\alpha} )\nn
\eea
we can get
\bea
\dfrac{1}{\sqrt{-g}}\dfrac{\delta \mathcal{L}}{\delta \Gamma_{\mu \nu}^{~~\lambda}}
&=&\nabla_\lambda g^{\mu\nu} - \Gamma_{\lambda\kappa}^{~~\mu} g^{\kappa\nu} - \Gamma_{\lambda\kappa}^{~~\nu} g^{\mu\kappa} - \delta_\lambda^\nu (\nabla_\rho g^{\mu\rho}- \Gamma_{\rho\kappa}^{~~\mu} g^{\kappa\rho} - \Gamma_{\rho\kappa}^{~~\rho} g^{\mu\kappa}) \nn \\
&&+\dfrac{1}{2}g^{\mu\nu} g^{\sigma\kappa}(\nabla_\lambda g_{\sigma\kappa}+ \Gamma_{\lambda\sigma}^{~~\alpha} g_{\alpha\kappa} + \Gamma_{\lambda\kappa}^{~~\alpha} g_{\sigma\alpha})\\
&&-\dfrac{1}{2}\delta^{\nu}_\lambda(\nabla^\mu g_{\sigma\kappa}+ \Gamma_{~\sigma}^{\mu~\alpha} g_{\alpha\kappa} + \Gamma_{~\kappa}^{\mu~\alpha} g_{\sigma\alpha}) g^{\sigma\kappa} \nn\\
&&-g^{\mu\nu}\Gamma_{\sigma\lambda}^{~~\sigma}
+ g^{\rho\mu}\Gamma_{\rho\lambda}^{~~\nu}-\delta_\lambda^\nu g^{\rho\sigma}\Gamma_{\sigma\rho}^{~~\mu}+g^{\nu\rho}\Gamma_{\lambda\rho}^{~~\mu} \nn \\
&=&0 \nn
\eea
and now let us define the non-metricity as $\nabla_\mu g_{\nu\lambda} \equiv - Q_{\mu\nu\lambda}$,\footnote{Note that $\nabla_\mu g^{\nu\lambda}=Q_\mu^{~\nu\lambda}$} and the torsion as $ t_{\mu\nu}^{~~\lambda}= \Gamma_{\mu\nu}^{~~\lambda} - \Gamma_{\nu\mu}^{~~\lambda}$. And let $\Gamma_{\mu\nu\lambda}= g_{\lambda\alpha} \Gamma_{\mu\nu}^{~~\alpha}$, $Q_{\alpha} = -\dfrac{1}{4}Q_{\alpha\kappa}^{~~\kappa}$. Then we have the following equation,
\be
t_{\lambda\mu}^{~~\nu}-\delta_\mu^\nu t_{\lambda\sigma}^{~~\sigma}+\delta_\lambda^\nu t_{\mu\sigma}^{~~\sigma}
-Q_{\lambda\mu}^{~~\nu}+\delta_\lambda^\nu Q_{\sigma\mu}^{~~\sigma}-2\delta_\mu^\nu Q_{\lambda}+2\delta_\lambda^\nu Q_{\mu}=0 \label{palatinieq}
\ee

If we treat the metric compatible connections, {\it i.e.} $Q_{\lambda\mu}^{~~\nu}=0$, then above equation (\ref{palatinieq}) becomes
\be
t_{\lambda\mu}^{~~\nu}-\delta_\mu^\nu t_{\lambda\sigma}^{~~\sigma}+\delta_\lambda^\nu t_{\mu\sigma}^{~~\sigma}=0,\label{palatinieq2}
\ee

and contracting this equation with $\delta_{\nu}^{\mu}$, we get
\be
t_{\lambda\sigma}^{~~\sigma}-4 t_{\lambda\sigma}^{~~\sigma}+ t_{\lambda\sigma}^{~~\sigma}=-2t_{\lambda\sigma}^{~~\sigma}=0,
\ee
again we put this result to eq.(\ref{palatinieq2}), then we have $t_{\lambda\mu}^{~~\nu}=0$.
So if this theory is metric compatible, it also should be torsion free.

Conversely, when $t_{\lambda\mu}^{~~\nu}=0$, from the eq.(\ref{palatinieq}) it is easy to get $Q_{\lambda\mu}^{~~\nu}=0$. Therefore torsion free connection should be metric compatible.

Now let us find the general relation between the torsion and the non-metricity.
First, after contracting eq.(\ref{palatinieq}) with $\delta_\nu^\mu$ and $\delta_\nu^\lambda$ respectively we can get
\be
-2 t_{\lambda\sigma}^{~~\sigma} + Q_{\sigma\lambda}^{~~\sigma} - 2 Q_{\lambda}=0 \label{1}
\ee
 and
\be
2 t_{\mu\sigma}^{~~\sigma} + 3 Q_{\sigma\mu}^{~~\sigma} +6 Q_{\mu}=0 \label{2}.
\ee

And adding both equations (\ref{1}) and (\ref{2}), we get $Q_{\sigma\mu}^{~~\sigma}=-Q_\mu$. And put this into eq.(\ref{2}), then we also get $t_{\mu\sigma}^{~~\sigma}= - \dfrac{3}{2} Q_\mu$. And again put these results into eq.(\ref{palatinieq}), then

\be
t_{\lambda\mu}^{~~\nu} - Q_{\lambda\mu}^{~~\nu} - \dfrac{1}{2} \delta_\lambda^\nu Q_\mu - \dfrac{1}{2} \delta_\mu^\nu Q_\lambda = 0 \label{4}.
\ee

And by symmetrizing and anti-symmetrizing eq.(\ref{4}) about the lower indices $\lambda$ and $\mu$, we get $Q_{(\lambda\mu)}^{~~\nu}= - Q_{(\lambda} \delta_{\mu)}^\nu$ and $Q_{[\lambda\mu]}^{~~\nu}= t_{\lambda\mu}^{~~\nu}$.

Therefore
\bea
Q_{\lambda\mu}^{~~\nu}&=&Q_{(\lambda\mu)}^{~~\nu}+Q_{[\lambda\mu]}^{~~\nu} \\ \nn
&=& - \dfrac{1}{2}(Q_{\lambda} \delta_{\mu}^\nu+Q_{\mu} \delta_{\lambda}^\nu )+t_{\lambda\mu}^{~~\nu},
\eea
and
\bea
Q_{\lambda\mu\nu}= - \dfrac{1}{2}Q_{\lambda} g_{\mu\nu} - \dfrac{1}{2}Q_{\mu}g_{\lambda\nu} +t_{\lambda\mu\nu}.\label{7}
\eea

Now the left hand side of eq.(\ref{7}) is $\mu\nu$-symmetric, so the right hand side also should be. So $- \dfrac{1}{2}Q_{\mu}g_{\lambda\nu} +t_{\lambda\mu\nu} \equiv A_{\lambda\mu\nu}=A_{\lambda(\mu\nu)}$, and

\be
t_{\lambda\mu\nu} = \dfrac{1}{2}Q_{\mu}g_{\lambda\nu}+A_{\lambda\mu\nu}. \label{8}
\ee

Now eq.(\ref{8}) should be $\lambda\mu$-antisymmetric so we can express eq.(\ref{8}) like following
\be
t_{\lambda\mu\nu} = \dfrac{1}{2}Q_{\mu}g_{\lambda\nu}-\dfrac{1}{2}Q_{\lambda}g_{\mu\nu}+B_{\lambda\mu\nu},
\ee
where $B_{\lambda\mu\nu}$ is some tensor which is symmetric about the second and third indices and antisymmetric about the first and second indices, but such a tensor cannot be exist. So $B_{\lambda\mu\nu}=0$.
Finally we have got the relation between the torsion and the non-metricity and it is following
\be
t_{\lambda\mu\nu} = \dfrac{1}{2}Q_{\mu}g_{\lambda\nu}-\dfrac{1}{2}Q_{\lambda}g_{\mu\nu}
\ee
and from eq.(\ref{7})
\bea
Q_{\lambda\mu\nu} &=& - \dfrac{1}{2}Q_{\lambda} g_{\mu\nu} - \dfrac{1}{2}Q_{\mu}g_{\lambda\nu}+\dfrac{1}{2}Q_{\mu}g_{\lambda\nu}-\dfrac{1}{2}Q_{\lambda}g_{\mu\nu} \nn \\
&=& - Q_\lambda g_{\mu\nu}\nn \\
\therefore \nabla_\lambda g_{\mu\nu} &=& Q_\lambda g_{\mu\nu} \label{10}
\eea

Now let us express the connection with the metric and Weyl vector field using eq.(\ref{10}), that is,
\bea
\partial_\mu g_{\nu\lambda} - \Gamma_{\mu\nu\lambda}-\Gamma_{\mu\lambda\nu} = Q_\mu g_{\nu\lambda} \nn \\
\therefore \Gamma_{\mu\nu\lambda}+\Gamma_{\mu\lambda\nu} = \partial_\mu g_{\nu\lambda} - Q_\mu g_{\nu\lambda}. \label{11}
\eea
By changing the order of indices, we get two expressions like following
\bea
\Gamma_{\nu\mu\lambda}+\Gamma_{\nu\lambda\mu} = \partial_\nu g_{\mu\lambda} - Q_\nu g_{\mu\lambda}, \label{12} \\
\Gamma_{\lambda\mu\nu}+\Gamma_{\lambda\nu\mu} = \partial_\lambda g_{\mu\nu} - Q_\lambda g_{\mu\nu}. \label{13}
\eea

And now adding (\ref{11}) and (\ref{12}) then subtracting (\ref{13}), and with some algebra we have the final expression
\bea
\Gamma_{\mu\nu\lambda} &=& \dfrac{1}{2}(\partial_\mu g_{\nu\lambda}+\partial_\nu g_{\mu\lambda}-\partial_\lambda g_{\mu\nu} ) \nn \\ &&- \dfrac{1}{2}( Q_\mu g_{\nu\lambda}+Q_\nu g_{\mu\lambda}-Q_\lambda g_{\mu\nu})+\dfrac{1}{2}(t_{\mu\nu\lambda}+t_{\lambda\mu\nu}+t_{\lambda\nu\mu}).\nn \\
&=& \dfrac{1}{2}(\partial_\mu g_{\nu\lambda}+\partial_\nu g_{\mu\lambda}-\partial_\lambda g_{\mu\nu} ) - \dfrac{1}{2} Q_\mu g_{\nu\lambda}.\label{connection ansatz}
\eea

By the same method we have got the same result (\ref{connection ansatz})  in a Lagrangian such as
\be
\mathcal{L}= \sqrt{-g}( a R + b \tilde{\epsilon}^{\alpha\beta\gamma\delta} R_{\alpha\beta\gamma\delta})
\ee

where $a$ and $b$ are some constants and $\tilde{\epsilon}_{\alpha\beta\gamma\delta} = \sqrt{-g}\epsilon_{\alpha\beta\gamma\delta}$, $\tilde{\epsilon}^{\alpha\beta\gamma\delta} = \dfrac{1}{\sqrt{-g}}\epsilon^{\alpha\beta\gamma\delta}$, $\epsilon_{0123}=1$, $\epsilon^{0123}=-1$.

In higher derivative gravity we cannot say that the above results is valid in general.
And it is not proper to apply the Palatini approach to higher derivative gravity\cite{buc79}.
So instead of applying this method to the higher derivative gravity,
we will just try to take eq.(\ref{connection ansatz}) as a constraint of connections and call it the Palatini connection. But at least we can say that in low energy regime with non-zero vacuum expectation value of the scalar field the Weyl vector fields can be the source of the torsion.

%------------------------------------------------------------------------
%\newpage
\subsection{General Lagrangian under the Palatini connections }
Since $Q_\mu = -f J W_\mu$ where $J(=2)$ is the weight of $g_{\mu\nu}$, we can express the torsion in terms of the Weyl vector fields. That is,
\bea
t_{\mu\nu}^{~~\lambda} &=& \dfrac{1}{2} ( \delta_\mu^\lambda Q_\nu - \delta_\nu^\lambda Q_\mu) = -\dfrac{1}{2} f J( \delta_\mu^\lambda W_\nu - \delta_\nu^\lambda W_\mu) \nn \\
&=& f( W_\mu \delta_\nu^\lambda  - W_\nu \delta_\mu^\lambda ).
\eea
This means that the Weyl vector fields generate the torsion fields in low energy regime.
And the Palatini connection (\ref{connection ansatz}) becomes
\bea
\Gamma_{\mu\nu\lambda} &=& \dfrac{1}{2}(\partial_\mu g_{\nu\lambda}+\partial_\nu g_{\mu\lambda}-\partial_\lambda g_{\mu\nu} ) + \dfrac{1}{2} f J W_\mu g_{\nu\lambda} \label{connection ansatz2} \\
&\equiv& \bar \Gamma_{\mu\nu\lambda} +  f W_\mu g_{\nu\lambda}.
\eea
Or
\bea
\Gamma_{\mu\nu}^{~~\lambda} &=& \bar \Gamma_{\mu\nu}^{~~\lambda} + f W_\mu \delta_{\nu}^{\lambda}.
\eea
The curvature tensor is
\bea
{R}_{\mu\nu\lambda}^{~~~\rho}&=&2\left\{ \partial_{[\mu}{\Gamma}_{\nu]\lambda}^{~~\rho}-{\Gamma}_{[\mu|\lambda|}^{~~\sigma} {\Gamma}_{\nu]\sigma}^{~~\rho}\right\} \nn \\
&=&2\left\{ \partial_{[\mu}\left(\bar{\Gamma}_{\nu]\lambda}^{~~\rho}+f W_{\nu]} \delta_{\lambda}^\rho \right)-\left(\bar{\Gamma}_{[\mu|\lambda|}^{~~\sigma}+f W_{[\mu} \delta_{|\lambda|}^\sigma \right)
\left( \bar {\Gamma}_{\nu]\sigma}^{~~\rho}+f W_{\nu]} \delta_{\sigma}^{\rho}\right) \right\} \nn \\
&=&2\left\{ \partial_{[\mu}\bar{\Gamma}_{\nu]\lambda}^{~~\rho}-\bar{\Gamma}_{[\mu|\lambda|}^{~~\sigma} \bar{\Gamma}_{\nu]\sigma}^{~~\rho}\right\} + 2f\left\{\delta_\lambda^\rho \partial_{[\mu}W_{\nu]}-\delta_\sigma^\rho W_{[\nu} \bar{\Gamma}_{\mu]\lambda}^{~~\sigma} -\delta_\lambda^\sigma W_{[\mu} \bar\Gamma_{\nu]\sigma}^{~~\rho} \right\}\nn \\
&=&2\left\{ \partial_{[\mu}\bar{\Gamma}_{\nu]\lambda}^{~~\rho}-\bar{\Gamma}_{[\mu|\lambda|}^{~~\sigma} \bar{\Gamma}_{\nu]\sigma}^{~~\rho}\right\} + 2f\delta_\lambda^\rho \partial_{[\mu}W_{\nu]} \\
&=&\bar R_{\mu\nu\lambda}^{~~~\rho} + 2f\delta_\lambda^\rho \partial_{[\mu}W_{\nu]} =\bar R_{\mu\nu\lambda}^{~~~\rho} + 2f\delta_\lambda^\rho \bna_{[\mu}W_{\nu]}. \nn
\eea
And the Ricci tensor and scalar are
\bea
R_{\mu\nu}&=&R_{\mu\sigma\nu}^{~~~\sigma} =\bar R_{\mu\nu} + 2f \bna_{[\mu}W_{\nu]} \\
R&=& R_\mu^{~\mu} = \bar R.
\eea
Since the Weyl-Cartan curvature does not have all the symmetry of the Riemann curvature, we have the another Ricci tensor which is defined by following
\bea
\check{R}_{\mu\nu}&\equiv&R_{\rho\mu\lambda\nu}g^{\rho\lambda}=(\bar R_{\rho\mu\lambda\nu} + 2f g_{\lambda\nu} \bna_{[\rho}W_{\mu]})g^{\rho\lambda} \nn \\
&=& \bar R_{\mu\nu}+2f \bna_{[\nu}W_{\mu]}=\bar R_{\mu\nu}-2f \bna_{[\mu}W_{\nu]}=R_{\nu\mu}
\eea
With these quantities we can compute the square of them. The squares of the curvature tensors are following
\bea
R_{\mu\nu\lambda\rho}R^{\mu\nu\lambda\rho}&=&(\bar R_{\mu\nu\lambda\rho} + 2f g_{\lambda\rho} \bna_{[\mu}W_{\nu]})(\bar R^{\mu\nu\lambda\rho} + 2f g^{\lambda\rho} \bna^{[\mu}W^{\nu]}) \nn \\
&=& \bar R_{\mu\nu\lambda\rho} \bar R^{\mu\nu\lambda\rho} + 16f^2 \bna_{[\mu}W_{\nu]}\bna^{[\mu}W^{\nu]} \\
R_{\mu\nu\lambda\rho}R^{\mu\nu\rho\lambda}&=& -\bar R_{\mu\nu\lambda\rho} \bar R^{\mu\nu\lambda\rho} + 16f^2 \bna_{[\mu}W_{\nu]}\bna^{[\mu}W^{\nu]} \\
R_{\mu\nu\lambda\rho}R^{\lambda\rho\mu\nu}&=&(\bar R_{\mu\nu\lambda\rho} + 2f g_{\lambda\rho} \bna_{[\mu}W_{\nu]})(\bar R^{\lambda\rho\mu\nu} + 2f g^{\mu\nu} \bna^{[\lambda}W^{\rho]}) \nn \\
&=&\bar R_{\mu\nu\lambda\rho} \bar R^{\lambda\rho\mu\nu}= \bar R_{\mu\nu\lambda\rho} \bar R^{\mu\nu\lambda\rho}
= -R_{\mu\nu\lambda\rho}R^{\lambda\rho\nu\mu} \\
R_{\mu\nu\lambda\rho}R^{\mu\lambda\nu\rho}&=&(\bar R_{\mu\nu\lambda\rho} + 2f g_{\lambda\rho} \bna_{[\mu}W_{\nu]})(\bar R^{\mu\lambda\nu\rho} + 2f g^{\nu\rho} \bna^{[\mu}W^{\lambda]})\nn \\
&=& \bar R_{\mu\nu\lambda\rho} \bar R^{\mu\lambda\nu\rho} + 4f^2 \bna_{[\mu}W_{\nu]}\bna^{[\mu}W^{\nu]}\nn \\
&=& -R_{\mu\nu\lambda\rho}R^{\nu\lambda\mu\rho}\\
R_{\mu\nu\lambda\rho}R^{\mu\lambda\rho\nu}&=&(\bar R_{\mu\nu\lambda\rho} + 2f g_{\lambda\rho} \bna_{[\mu}W_{\nu]})(-\bar R^{\mu\lambda\nu\rho} + 2f g^{\nu\rho} \bna^{[\mu}W^{\lambda]})\nn \\
&=& -\bar R_{\mu\nu\lambda\rho} \bar R^{\mu\lambda\nu\rho} + 4f^2 \bna_{[\mu}W_{\nu]}\bna^{[\mu}W^{\nu]}\\
R_{\mu\nu\lambda\rho}R^{\nu\rho\lambda\mu}&=&(\bar R_{\mu\nu\lambda\rho} + 2f g_{\lambda\rho} \bna_{[\mu}W_{\nu]})(-\bar R^{\mu\lambda\nu\rho} + 2f g^{\lambda\mu} \bna^{[\nu}W^{\rho]})\nn \\
&=& -\bar R_{\mu\nu\lambda\rho} \bar R^{\mu\lambda\nu\rho} - 4f^2 \bna_{[\mu}W_{\nu]}\bna^{[\mu}W^{\nu]}\nn \\
&=&-R_{\mu\nu\lambda\rho}R^{\mu\lambda\nu\rho} = -R_{\mu\nu\lambda\rho}R^{\rho\nu\lambda\mu}\\
%R_{\mu\nu\lambda\rho}R^{\nu\lambda\mu\rho}&=&\bar R_{\mu\nu\lambda\rho} \bar R^{\nu\lambda\mu\rho} - 4f^2 \bna_{[\mu}W_{\nu]}\bna^{[\mu}W^{\nu]}\\
R_{\mu\nu\lambda\rho}R^{\rho\lambda\nu\mu}&=&\bar R_{\mu\nu\lambda\rho} \bar R^{\mu\nu\lambda\rho}=R_{\mu\nu\lambda\rho}R^{\lambda\rho\mu\nu}\\
R_{\mu\nu\lambda\rho}R^{\rho\nu\mu\lambda}&=&-\bar R_{\mu\nu\lambda\rho} \bar R^{\mu\lambda\nu\rho} + 4f^2 \bna_{[\mu}W_{\nu]}\bna^{[\mu}W^{\nu]}= R_{\mu\nu\lambda\rho} R^{\mu\lambda\rho\nu}.
\eea
And the square of the Ricci tensors and scalars are following
\bea
R_{\mu\nu}R^{\mu\nu}&=&(\bar R_{\mu\nu} + 2f \bna_{[\mu}W_{\nu]})(\bar R^{\mu\nu} + 2f \bna^{[\mu}W^{\nu]}) \nn \\
&=&\bar R_{\mu\nu}\bar R^{\mu\nu} +4f^2 \bna_{[\mu}W_{\nu]}\bna^{[\mu}W^{\nu]} \\
R_{\mu\nu}R^{\nu\mu} &=& \bar R_{\mu\nu}\bar R^{\mu\nu}-4f^2 \bna_{[\mu}W_{\nu]}\bna^{[\mu}W^{\nu]} \\
R^2 &=& \bar R^2.
\eea
Now let us consider the Gauss-Bonnet like identity.
\bea
I_{BF} &\equiv& R^2 - (R_{\mu\nu}+ \check R_{\mu\nu})(R^{\nu\mu}+ \check R^{\nu\mu})+R_{\mu\nu\alpha\beta} R^{\alpha\beta\mu\nu}.
\eea
But in our case it is the same with the Gauss-Bonnet identity of Riemannian. That is,
\bea
I_{BF} &\equiv& R^2 - (R_{\mu\nu}+\check R_{\mu\nu})(R^{\nu\mu}+\check R^{\nu\mu})+R_{\mu\nu\alpha\beta} R^{\alpha\beta\mu\nu} \nn \\
&=& \bar R^2 - (\bar R_{\mu\nu} + 2f \bna_{[\mu}W_{\nu]}+\bar R_{\mu\nu} - 2f \bna_{[\mu}W_{\nu]})(\bar R^{\nu\mu} + 2f \bna^{[\nu}W^{\mu]}+\bar R^{\nu\mu}\nn\\ &&- 2f \bna^{[\nu}W^{\mu]})
+ \bar R_{\mu\nu\alpha\beta} \bar R^{\mu\nu\alpha\beta} \nn \\
&=& \bar R^2 + 4 \bar R_{\mu\nu} \bar R^{\mu\nu}-\bar R_{\mu\nu\alpha\beta} \bar R^{\mu\nu\alpha\beta}= I_{GB}(\bar R)
\eea
So the topological invariant Lagrangian has no dynamics of the Weyl vector fields. Now let us think of the general type Lagrangian which is similar to the Lagrangian in \cite{generaltorsion}.
\bea
\mathcal{L}_{gen}&=& a_0 R_{\mu\nu\lambda\kappa}R^{\mu\nu\lambda\kappa}+a_1 R_{\mu\nu\lambda\kappa} R^{\lambda\kappa\mu\nu}+ a_2 R_{\mu\nu\lambda\kappa}R^{\mu\lambda\nu\kappa}+a_3 R_{\mu\nu\lambda\kappa}R^{\mu\nu\kappa\lambda}\nn \\
&&+ a_4 R_{\mu\nu\lambda\kappa}R^{\mu\lambda\kappa\nu}  + a'_5 R_{\mu\nu}R^{\mu\nu} + a'_6 R_{\mu\nu}R^{\nu\mu} + a'_7 \check{R}_{\mu\nu}\check{R}^{\mu\nu} \nn \\
&& + a'_8 R_{\mu\nu}\check{R}^{\mu\nu} + a'_9 \check{R}_{\mu\nu}\check{R}^{\nu\mu} + a'_{10} R_{\mu\nu}\check{R}^{\nu\mu}+ a_7 R^2+a_{8} A_{\mu\nu\lambda\kappa}A^{\mu\nu\lambda\kappa}\nn \\
&=& a_0 R_{\mu\nu\lambda\kappa}R^{\mu\nu\lambda\kappa}+a_1 R_{\mu\nu\lambda\kappa} R^{\lambda\kappa\mu\nu}+ a_2 R_{\mu\nu\lambda\kappa}R^{\mu\lambda\nu\kappa}+a_3 R_{\mu\nu\lambda\kappa}R^{\mu\nu\kappa\lambda} \nn \\
&&+ a_4 R_{\mu\nu\lambda\kappa}R^{\mu\lambda\kappa\nu} + (a'_5 +a'_7 +a'_{10}) R_{\mu\nu}R^{\mu\nu}  \\
&&+ (a'_6 + a'_8 + a'_9) R_{\mu\nu}R^{\nu\mu} +a_{7}R^2+a_{8} A_{\mu\nu\lambda\kappa}A^{\mu\nu\lambda\kappa}\nn
\eea
where $A_{\mu\nu\lambda\kappa}= \dfrac{1}{6}(R_{\mu\nu\lambda\kappa}+R_{\mu\lambda\kappa\nu}+R_{\mu\kappa\nu\lambda}+R_{\lambda\kappa\mu\nu}+R_{\nu\lambda\kappa\mu}+R_{\nu\lambda\mu\kappa})$ which vanishes in Riemann space-time. After redefining the coefficients, we can write the Lagrangian like following
\bea
\mathcal{L}_{gen}
&=& a_0 R_{\mu\nu\lambda\kappa}R^{\mu\nu\lambda\kappa}+a_1 R_{\mu\nu\lambda\kappa} R^{\lambda\kappa\mu\nu}+ a_2 R_{\mu\nu\lambda\kappa}R^{\mu\lambda\nu\kappa}+a_3 R_{\mu\nu\lambda\kappa}R^{\mu\nu\kappa\lambda} \nn \\
&&+ a_4 R_{\mu\nu\lambda\kappa}R^{\mu\lambda\kappa\nu} + a_5 R_{\mu\nu}R^{\mu\nu} + a_6 R_{\mu\nu}R^{\nu\mu} +a_{7}R^2+a_{8} A_{\mu\nu\lambda\kappa}A^{\mu\nu\lambda\kappa} \nn \\
&=&(a_0 +a_1 -a_3 + \dfrac{1}{3} a_8 ) \bar R_{\mu\nu\lambda\kappa} \bar R^{\mu\nu\lambda\kappa} + (a_2 -a_4  + \dfrac{7}{18} a_8 ) \bar R_{\mu\nu\lambda\kappa} \bar R^{\mu\lambda\nu\kappa} \nn \\
&&+(a_5+a_6)\bar R_{\mu\nu} \bar R^{\mu\nu}+ a_7 \bar R^2  \\
&& + 4(4a_0+a_2+4a_3+a_4+a_5-a_6+\dfrac{8}{9}a_8)f^2 \bna_{[\mu}W_{\nu]}\bna^{[\mu}W^{\nu]} \nn \\
&=&(a_0 +a_1 + \dfrac{1}{2} a_2-a_3 - \dfrac{1}{2} a_4 + \dfrac{19}{36} a_8 ) \bar R_{\mu\nu\lambda\kappa} \bar R^{\mu\nu\lambda\kappa}  \nn \\
&&+(a_5+a_6)\bar R_{\mu\nu} \bar R^{\mu\nu}+ a_7 \bar R^2  \\
&& + 4(4a_0+a_2+4a_3+a_4+a_5-a_6+\dfrac{8}{9}a_8)f^2 \bna_{[\mu}W_{\nu]}\bna^{[\mu}W^{\nu]} \nn
\eea
where we have used following
\bea
0&=&\bar R^{\mu\nu\lambda\kappa} ( \bar R_{\mu\nu\lambda\kappa} +\bar R_{\mu\lambda\kappa\nu} +\bar R_{\mu\kappa\nu\lambda}) \nn \\
&=& \bar R^{\mu\nu\lambda\kappa} ( \bar R_{\mu\nu\lambda\kappa} -\bar R_{\mu\lambda\nu\kappa} -\bar R_{\mu\lambda\nu\kappa}) \nn \\
&=&\bar R^{\mu\nu\lambda\kappa} ( \bar R_{\mu\nu\lambda\kappa} - 2\bar R_{\mu\lambda\nu\kappa}).
\eea
Thus if we put some constants into the coefficients, we can get some special Lagrangian such as the Yang-Mills type Lagrangian by setting $a_0 = - \frac{1}{4}$, $a_i =0$ ($i=1,~2,~\cdots, ~8$).
But we will keep the general form and for simplicity we want to write the Lagrangian as
\bea
\mathcal{L}_{gen}
&=&\alpha \bar R_{\mu\nu\lambda\kappa} \bar R^{\mu\nu\lambda\kappa} +\beta \bar R_{\mu\nu} \bar R^{\mu\nu}+ a_7 \bar R^2 + \gamma \bna_{[\mu}W_{\nu]}\bna^{[\mu}W^{\nu]}\label{generalweylLag}
\eea
where $\alpha=a_0 +a_1 + \frac{1}{2} a_2-a_3 - \frac{1}{2} a_4 + \frac{19}{36} a_8 $, $\beta=a_5+a_6$
and $\gamma=4f^2(4a_0+a_2+4a_3+a_4+a_5-a_6+\frac{8}{9}a_8)$.

Now $(\bar \nabla_{[\mu}W_{\nu]})^2= \frac{1}{2}(\bar \nabla_\mu W_\nu )^2  - \frac{1}{2} \bar R_{\mu\nu} W^\mu W^\nu - \frac{1}{2}\bar \nabla_\mu (W^\nu \bar \nabla_\nu W^\mu ) + \frac{1}{2} W^\nu \bar \nabla_\nu \bar \nabla_\mu W^\mu$, so we can express the Lagrangian like following
\bea
\mathcal{L}_{gen}
&=&\alpha \bar R_{\mu\nu\lambda\kappa} \bar R^{\mu\nu\lambda\kappa} +\beta \bar R_{\mu\nu} \bar R^{\mu\nu}+ a_7 \bar R^2 \\ &&+ \dfrac{\gamma}{2} \Big \{(\bar \nabla_\mu W_\nu )^2  -  \bar R_{\mu\nu} W^\mu W^\nu - \bar \nabla_\mu (W^\nu \bar \nabla_\nu W^\mu ) +  W^\nu \bar \nabla_\nu \bar \nabla_\mu W^\mu \Big \}. \nn
\eea

If we want the Lagrangian which reduces to Gauss-Bonnet invariant in the limit of Riemannian geometry, then we have only to put $\alpha = a_7$ and $\beta=-4a_7$.

Note that the Lagrangian (\ref{generalweylLag}) is just the form of the Maxwell theory in the curved space-time. \textit{So we can say that in low energy regime the Palatini connection make the Weyl symmetry broken but we have another gauge symmetry of $U(1)$}, that is, $\delta W_\mu = \pro_\mu \Lambda'$ and $\delta g_{\mu\nu}=0$. The Weyl fields and torsion fields are the geometric fields. And by some symmetry breaking we have got the Maxwell fields from the geometric fields. So this can be another type of the unification. Note that in \cite{generaltorsion} we showed the contortion field has $U(1)$ symmetry. Maybe there is a relation between the two $U(1)$ symmetry of contortion and Weyl vector fields. If we think this two $U(1)$ is identical and we set $K_\mu = 3 f W_\mu$, then the total connection becomes
\bea
\Gamma_{\mu\nu\lambda}&=& \bar \Gamma _{\mu\nu\lambda} + K_{\mu\nu\lambda}+f(g_{\mu\lambda}W_\nu+g_{\nu\lambda}W_\mu-g_{\mu\nu}W_\lambda) \nn \\
&=&\dfrac{1}{2}(\partial_\mu g_{\nu\lambda}+\partial_\nu g_{\mu\lambda}-\partial_\lambda g_{\mu\nu} ) +M_{\mu\nu\lambda} + \dfrac{1}{3}(g_{\mu\nu}K_\lambda - g_{\mu\lambda}K_\nu )\nn \\
&&+\dfrac{1}{6}\tilde \epsilon _{\mu\nu\lambda\kappa}S^\kappa +f(g_{\mu\lambda}W_\nu+g_{\nu\lambda}W_\mu-g_{\mu\nu}W_\lambda) \nn \\
&=&\dfrac{1}{2}(\partial_\mu g_{\nu\lambda}+\partial_\nu g_{\mu\lambda}-\partial_\lambda g_{\mu\nu} ) +M_{\mu\nu\lambda} +\dfrac{1}{6}\tilde \epsilon _{\mu\nu\lambda\kappa}S^\kappa +f W_\mu g_{\nu\lambda} \nn
\eea
where $K_\mu = K^{\sigma}_{~\sigma \mu}$, $S^\mu = \tilde \epsilon ^{\mu\nu\lambda\kappa} K_{\nu\lambda\kappa} $, $M^{\sigma}_{~\sigma \mu}=0$ and $\tilde \epsilon ^{\mu\nu\lambda\kappa} M_{\nu\lambda\kappa}=0$. So with this connection we can construct the theory of torsion and Weyl vector fields which has $U(1)$ symmetry. Note that in this case it is not that the Weyl gauge symmetry changes into $U(1)$ symmetry. The $U(1)$ symmetry comes from the torsion's symmetry. That is, we can say that the Weyl vector fields eat the torsion and become the Maxwell vector fields.

By the way there may be someone who wants to change the scale symmetry to the phase symmetry directly, then he should expand the real Weyl gauge transformation to the complex Weyl gauge transformation. That is, the vielbein transform like
\bea
{e'}^{~a}_\mu &=& e^{\Lambda(x)+ i \phi(x)} e^{~a}_\mu \\
{e'}_{~a}^\mu &=& e^{-\Lambda(x)+i \phi(x)} e_{~a}^\mu
\eea
where the Roman alphabet letters ($a, b, c, \cdots$) indicate the Lorentz index and the Greek alphabet letters ($\mu, \nu, \lambda, \cdots$) indicate the coordinate index.

And the metric can be induced from vielbein like following way,
\bea
g_{\mu\nu}&=& 2 {e^*}^{~a}_{(\mu} e_{\nu) a} \\
\eta_{ab} &=& 2 {e^*}_{~(a}^{\mu} e_{|\mu| b)}
\eea
Then the metric transforms like same way of the real Weyl transformation : $g'_{\mu\nu}=e^{2\Lambda(x)}g_{\mu\nu}$.
Now we can give weight to the tensor. The weight $(J_1, J_2)$ tensors transform like
$T'(x) = e^{J_1 \Lambda(x)+i J_2 \phi(x)}T(x)$. So the vielbeins $e^{~a}_\mu$ have the weight $(1, 1)$ and $e_{~a}^\mu$ have the weight $(-1, 1)$.
In this case the covariant derivative becomes
\bea
D_\mu = \partial_\mu - \Gamma_\mu + f_1 J_1 W_{1\mu} + f_2 J_2 W_{2\mu}
\eea
where the two types of Weyl vector fields transform like following,
\bea
W'_{1\mu}&=&W_{1\mu}- \dfrac{1}{f_1}\partial_\mu \Lambda \\
W'_{2\mu}&=&W_{2\mu}- \dfrac{i}{f_2}\partial_\mu \phi
\eea
In this way he may obtain the Maxwell fields from the Weyl vector fields. But this is a another story.
So we don't go further here, and leave it for another study.

%====================================================================
\section{Conclusions}\label{conclusion}

 The Weyl vector fields also can play an important role in quantum gravity with the torsion. And in low energy regime with scalar field there is a relation between the Weyl vector fields and the torsion fields. If this condition is given to Weyl vector fields and torsions, then the Lagrangian becomes like Maxwell type. And this Maxwell symmetry comes from the symmetry of torsion, not from the Weyl gauge symmetry.

%{\bf Acknowledgements}

\end{document}